\documentclass[prl,twocolumn,aps,a4paper,superscriptaddress,showpacs]{revtex4}
\usepackage{amsmath}
\usepackage[dvips]{graphicx}
\begin{document}

\newcommand\beq{\begin{equation}}
\newcommand\eeq{\end{equation}}
\newcommand\bea{\begin{eqnarray}}
\newcommand\eea{\end{eqnarray}}
\newcommand\bseq{\begin{subequations}} 
\newcommand\eseq{\end{subequations}}
\newcommand\bcas{\begin{cases}}
\newcommand\ecas{\end{cases}}

\title{Frame-independence of the Inhomogeneous Mixmaster Chaos via Misner-Chitr\'e-like variables}

\author{Riccardo Benini}
\email{riccardobenini@virgilio.it}
\affiliation{Dipartimento di Fisica - Universit\`a di Bologna and INFN - Sezione di Bologna,
via Irnerio 46, 40126 Bologna, Italy}
\author{Giovanni Montani}
\email{montani@icra.it} 
\affiliation{Dipartimento di Fisica Universit\`a di Roma ``La Sapienza''}
\affiliation{ICRA---International Center for Relativistic Astrophysics  
c/o Dipartimento di Fisica (G9) Universit\`a di Roma ``La Sapienza'',
Piazza A.Moro 5 00185 Rome, Italy}

\today

\begin{abstract}

We outline the covariant nature,with respect to the choice of a reference frame, of the chaos characterizing the generic cosmological solution near the initial singularity, i.e. the so-called inhomogeneous Mixmaster model.\
Our analysis is based on a "gauge" independent ADM-reduction of the dynamics to the physical degrees of freedom.\
The out coming picture shows how the inhomogeneous Mixmaster model is isomorphic point by point in space to a billiard on a Lobachevsky plane. Indeed, the existence of an asymptotic (energy-like) constant of the motion allows to construct the Jacobi metric associated to the geodesic flow and to calculate a non-zero Lyapunov exponent in each space point.\
The chaos covariance emerges from the independence of our scheme with respect to the form of the lapse function and the shift vector; the origin of this result relies on the dynamical decoupling of the space-points which takes place near the singularity, due to the asymptotic approach of the potential term to infinite walls.
At the ground of the obtained dynamical scheme is the choice of Misner-Chitr\' like variables which allows to fix the billiard potential walls.

\end{abstract}



\maketitle

\section{Basic Statements}

The homogeneous and isotropic Friedmann-Lemaitre-Robertson-Walker
 (FLRW) metric provides a valuable framework to describe
  the history of the Universe up to its very early stages of evolution.\
Indeed the very good agreement between the light elements abundances
 predicted for the primordial nucleosynthesis and the observed one allows to 
 extrapolate backward in time the FLRW dynamics up to 
 $10^{-2}-10^{-3}$ s \cite{KT90}; furthermore, recent 
 observations of the Cosmic Microwave Background (CMB) radiation 
  \cite{deB00,MAP}, suggest that an inflationary scenario took
 place and that the Universe was homogeneous and isotropic
 on the horizon scale up to $\mathcal{O}(10^{-32})s$.\
In spite of such an experimental evidence in favor of the FLRW model, there are mainly two well-grounded reasons to believe that, when the Universe temperature was above the Grand Unification scale ($\mathcal{O}(10^{15}$ GeV)) and below the Planck mass ($\mathcal{O}(10^{19}$ GeV)), the Universe was appropriately described by a generic inhomogeneous cosmological model \cite{BKL82,K93,M95}; indeed we have to stress that:
\begin{enumerate}
\item {as shown in \cite{KL63} the Universe in an expanding picture is stable with respect to tensorial perturbations; in fact the amplitude of a gravitational wave decays like the inverse of the scale factor \cite{LL,GRI}.
Therefore reversing the expanding behavior into a collapsing one, the homogeneity and the isotropy of the Universe become unstable with respect to wave like perturbations. In particular in \cite{GRI} is shown how a Bianchi type IX model far from the singularity can be represented (in term of exact solution) by a closed FLRW model plus small gravitational ripples; instead, close to the singularity this model has a fully developed anisotropy which results into a chaotic behavior. This issue is valuable because the Bianchi type IX cosmology presents features extensible (point by point in space) to the Generic Cosmological Solution}.
\item{It is commonly believed \cite{HH83, KT90} that the actual Universe came out from a quantum regime which was fully developed during the Planckian era and approached a classical limit (see \cite{KM97J}) only in a later stage of evolution.\
Since the early Universe contained more than a single horizon (see also \cite{M03A}), then during the Planckian era the metric and topology quantum fluctuations had to take place independently over two causally disconnected regions, and hence the survival of any global symmetry is prevented. In view of this, we infer that the quantum behavior of the early Universe has to be properly analyzed only in terms of a generic inhomogeneous model.}
\end{enumerate}
With respect to these two points, it is worth noting that the bridge between the generic evolution and the FLRW dynamics is provided on the horizon scale just by an inflationary scenario, as outlined in \cite{KM02}.

As well known, the generic cosmological solution \cite{BKL82} is characterized near the Big-Bang by a chaotic evolution which reduces the space-time to the structure of a foam \cite{K93, M95}.\
When approaching the cosmological singularity, the space points dynamically decouple and a time evolution resembling the so-called Mixmaster dynamics of the Bianchi VIII and IX models takes place independently in each of them \cite{BKL70,M69}(for further discussions on inhomogeneous cosmological models see \cite{WIB,BERGER}); from a physical point of view, space neighborhood is here considered at horizon size (for recent discussions on Mixmaster covariance see\cite{M01,M03,CL97}).

In recent years, the chaoticity of the homogeneous Mixmaster model has been widely studied in the literature (see \cite{H94,KM97P,CL97,IM01,M01,M03}) in view of understanding the features is of its covariant nature.\\
 Two convincing arguments, appeared in \cite{CL97, IM01}, support the idea that the Mixmaster chaos (described by the invariant measure introduced in \cite{CB83, KM97P}) remains valid in any system of coordinates.\\
 The main issue of the present work is to show that the property of space-time covariance can be extended to the inhomogeneous Mixmaster model.
In Section 2 we provide a gauge independent analysis (i.e. independent of the choice of the lapse function as well as of the shift vector) for the dynamics of the gravitational degrees of freedom. In Section 3 we discuss the asymptotic behavior of the potential term associated with the Ricci 3-scalar, showing how it can be modeled in terms of a potential wall surrounding a (dynamically) closed domain.
In Section 4 in view of the existence of an energy-like constant of motion, we construct the Jacobi metric associated with the resulting billiard on a two dimensional Lobachevsky plane point by point in space, and finally we calculate the covariant non-zero Lyapunov exponent.
The concluding remarks presented in Section 5 are devoted to outline how the approximation used for the potential term is completely self-consistent being dynamically induced. It is worth noting how we show that for any choice of lapse function and the shift-vector, taking Misner-Chitr\'e like variables, the system results to be chaotic; however, in this sense, our analysis states the independence of the chaos on the choice of the space-time coordinates, but the questions of covariance remains open when using different configuration coordinates because the Lyapunov exponent is not a good indicator for all of them.

\section{Hamiltonian Formulation}

A generic cosmological solution is represented by a gravitational field having available all its degrees of 
freedom and, therefore, allowing to specify a generic Cauchy problem.\\
In the Arnowitt-Deser-Misner (ADM) formalism, the metric tensor
corresponding to such a generic model takes the form
\begin{equation}
	d\Gamma^2=N^2 dt^2-\gamma_{\alpha\beta}(dx^\alpha+N^\alpha dt)(dx^\beta+N^\beta dt)
\end{equation}
where $N$ and $N^\alpha$ denote (respectively) the lapse function and the shift-vector, $\gamma_{\alpha\beta}$ ($\alpha,\beta=1,2,3$) the 3-metric tensor of the spatial hyper-surfaces $\Sigma^3$ for which $t=const$, being
\begin{equation}
\label{parametrizzazione della metrica}
	\gamma_{\alpha\beta}=e^{q_a}\delta_{ad}O^a_b O^d_c \partial_\alpha y^b \partial_\beta y^c,\ \ \ 
	a,b,c,d,\alpha,\beta=1,2,3,
\end{equation} 
while $q^a=q^a(x,t)$ and $y^b=y^b(x,t)$ six scalar functions, and $O^a_b=O^a_b(x)$ a $SO(3)$ matrix.\
By the metric tensor (\ref{parametrizzazione della metrica}), the action for the gravitational field is
\begin{equation}
\label{azione standard}
	S=\int_{\Sigma^{(3)}\times\Re}dt d^3 x\left(p_a\partial_t q^a+\Pi_d\partial_t y^d -NH-N^\alpha H_\alpha\right)\,,
\end{equation}
where
\begin{equation}	
	\label{vincoli hamiltoniani} 
	H=\frac{1}{ \sqrt \gamma}[\sum_a (p_a)^2-\frac{1}{2}(\sum_b p_b)^2-\gamma ^{(3)}R]
\end{equation}	
\begin{equation}
\label{vincoli hamiltoniani2}	 
	 H_\alpha=\Pi_c \partial_\alpha y^c +p_a \partial_\alpha q^a +2p_a(O^{-1})^b_a\partial_\alpha O^a_b;
\end{equation}
in (\ref{vincoli hamiltoniani}) and (\ref{vincoli hamiltoniani2}) $p_a$ and $\Pi_d$ are the conjugate momenta of the variable $q^a$ and $y^b$ respectively, and the $^{(3)}R$ is the Ricci 3-scalar which plays the role of a potential term.\\

The ten independent components of a generic metric tensor are represented by the three scale factors $q^a$, the three degrees of freedom $y^a$, the lapse function $N$ and the three components of the shift-vector $N^a$; it is worth noting that, by the variation of the variables $p_a$, $\Pi_a$ in the action (\ref{azione standard}), the relations:
\begin{equation}
	\label {condizione di gauge sullo shift-vector}
	\partial_t y^d=N^\alpha\partial_\alpha y^d
\end{equation}
\begin{equation}
  \label {condizione di gauge sulla lapse function}
	N=\frac{\sqrt{\gamma}}{\sum_a p_a}\left(N^\alpha\partial_\alpha\sum_b q^b-\partial_t\sum_b q^b\right).
\end{equation}
take place.

\section {ADM-reduction of the dynamics}

We use the Hamiltonian constraints $H=H_\alpha=0$ for the reduction of the dynamics to the physical degrees of freedom; from (\ref{vincoli hamiltoniani}) and (\ref{vincoli hamiltoniani2}), we note that the super-momentum constraints can be diagonalized and explicitly solved by choosing the function $y^a$ as special coordinates, i.e. taking the transformation $\eta=t$, and $y^a=y^a(t,x)$; in fact starting by (\ref{vincoli hamiltoniani}) and (\ref{vincoli hamiltoniani2}) we get the expression 
\begin{equation}
	\Pi_b=-p_a\frac{\partial q^a}{\partial y^b}-2p_a(O^{-1})^c_a\frac{\partial O^a_c}{ \partial y^b}\,.
\end{equation}
Furthermore, in the new coordinates we have
\begin{equation}
	\begin{cases}
	q^a(t,x)\to q^a(\eta,y)\cr
	p_a(t,x)\to p'_a(\eta,y)={p_a(\eta,y)/|J|}\cr
	\frac{\partial}{ \partial t}\rightarrow \frac{\partial y^b}{ \partial t}\frac{\partial}{ \partial y^b}+\frac{\partial}{ \partial \eta}\cr
	\frac{\partial}{ \partial x^\alpha}\rightarrow \frac{\partial y^b}{ \partial x^\alpha}\frac{\partial}{ \partial y^b}\cr
	\end{cases}
	\,,
\end{equation}
where $|J|$ denotes the Jacobian of the transformation.\
The first relation holds in general for all the scalar quantities, while the second one for all the scalar densities;
hence the action (\ref{azione standard}) rewrites as
\begin{equation}
	\label{finale non approssimata}
	S=\int_{\Sigma^{(3)}\times\Re}d\eta d^3 y \left(p_a\partial_\eta q^a+2p_a(O^{-1})^c_a\partial_\eta O^a_c-NH\right)\,.
\end{equation} 

\section{The potential wall and the reduced variational principle}

\begin{figure} 

\begin{center} 

\includegraphics[width=\hsize]{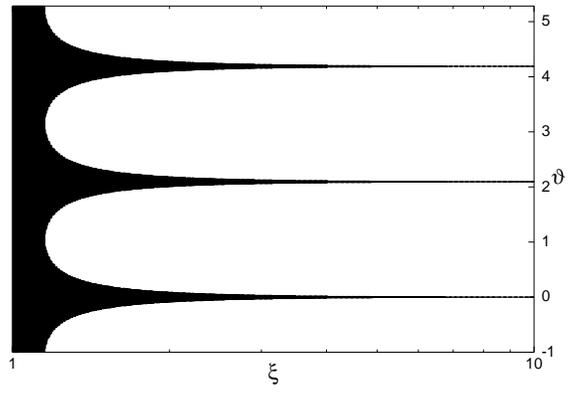}

\caption{ $\Gamma_Q(\xi,\theta)$\label{fig:cuspidi2 cap5}} 


\end{center}

\end{figure}

The potential term appearing in the super-Hamiltonian reads, in obvious notation, as:
\begin{equation}
\label{pot}
	U=\frac{D }{ |J|^2}\ ^{(3)}R=\sum_a \lambda_a^2 D^{2 Q_a}+\sum_{b\neq c}D^{Q_b+Q_c}\mathcal{O}\left(\partial q, (\partial q)^2,y,\eta\right)
\end{equation}
where 
\begin{equation}
	D\equiv \exp{\sum_a q^a},
\end{equation}
\begin{equation}
\label{anisotropia}
	Q_a\equiv \frac{q^a}{ \sum_b q^b}
\end{equation}
\begin{equation}
	\lambda_a^2\equiv \sum_{kj}\left(O^a_b \vec\nabla O^a_c (\vec\nabla y^c\wedge	\vec\nabla y^b)^2\right)\, ;
\end{equation}
Assuming the functions $y^a(t,x)$ smooth enough (which implies by (\ref{condizione di gauge sullo shift-vector}),(\ref{condizione di gauge sulla lapse function}) that the coordinates system is smooth "itself"), then all the gradients appearing in the potential $U$ are regular. Indeed this notion of regularity is not to be intended in absolute sense; in fact what really matter here is not that the gradient increase but simply that their behavior is not so strongly divergent to destroy the billiard representation (see next paragraph). In \cite{K93} it was shown that the spatial gradients increase logarithmically in the proper time along the billiard's geodesic and therefore result to be of higher order.
Thus, as $D\to 0$ \footnote{It's worth noting that the quantity D is proportional to the determinant of the three-metric, being $\det(\gamma_{\alpha\beta})=|J|\exp{\sum_a q^a}=|J|D$, and controlls it's vanishing behavior} the spatial curvature $^{(3)}R$ diverges and the cosmological singularity appears; in this limit, the first term of the potential $U$ dominates all the remaining ones and can be modeled by the potential wall
\begin{equation}
\label{U}
	U=\sum_a{\Theta(Q_a)}\, ,
\end{equation}
being
\begin{equation}
	\label{funzioni generalizzate}
	\Theta(x)=\begin{cases}
	+\infty\ &  $if$\  x>0,\cr 0\ &  $if$\  x<0.\cr
	\end{cases}
\end{equation}
By (\ref{U}) the Universe dynamics evolves independently in each space point; the point-Universe can move only within a dynamically-closed domain $\Gamma_Q$ (see figure (\ref{fig:cuspidi2 cap5})) (indeed the three corners at infinity are open but they correspond to a set of zero measure in the space of initial conditions).
Since in $\Gamma_Q$ the potential $U$ asymptotically vanishes, near the singularity we have $\partial p_a/ \partial\eta=0$; and then the term $2p_a(O^{-1})^c_a\partial_\eta O^a_c$ in (\ref{finale non approssimata}) behaves as an exact time-derivative and can be ruled out of the variational principle. 
The ADM reduction is completed by introducing the so-called Misner-Chitr\'e like variables \cite{C72, K93,M00, IM01} as 
\begin{equation}
	\label{cambio di chitre}
	\begin{cases}
	q^l=e^\tau [\sqrt{\xi^2-1}(\cos\theta+\sqrt 3 \sin\theta)-\xi]\cr
	q^m=e^\tau [\sqrt{\xi^2-1}(\cos\theta-\sqrt 3 \sin\theta)-\xi]\cr
	q^n=-e^\tau(\xi+2\sqrt{\xi^2-1}\cos\theta)
	\end{cases} 
\end{equation}

The way in which the anisotropy parameters $Q_a$ (\ref{anisotropia}) are rewritten is an  important feature of these variables since we easily get the $\tau$-independent expressions:
\[
Q_1=\frac{1}{ 3}-\frac{\sqrt{\xi^{2}-1}}{ 3\xi}(\cos\theta+\sqrt{3}\sin\theta)
\]
\begin{equation}
\label{parametri di anisotropia32}
        Q_2=\frac{1}{ 3}-\frac{\sqrt{\xi^{2}-1}}{ 3\xi}(\cos\theta-\sqrt{3}\sin\theta)
\end{equation}
\[
Q_3=\frac{1}{ 3}+\frac{2\sqrt{\xi^{2}-1}}{ 3\xi}\cos\theta
\]

When expressed in term of such variables the super-hamiltonian constraint can be solved in the domain $\Gamma_Q$:
\begin{equation}
	\label{hamiltoniana ADM}
	-p_\tau\equiv\epsilon=\sqrt{(\xi^2-1)p_\xi^2+\frac{p_\theta^2}{\xi^2-1}}
\end{equation}
and the reduced action reads as
\begin{equation}
	\label{hamiltoniana ridotta}
	\delta S_{\Gamma_Q}=\delta\int d\eta d^3 y (p_\xi\partial_\eta\xi+p_\theta\partial_\eta\theta-\epsilon\partial_\eta\tau)=0\,.
\end{equation}
By the asymptotic limit (\ref{U}) and the Hamilton equations associated with (\ref{hamiltoniana ridotta}) it follows that $\epsilon$ is a constant of motion, i.e. $d\epsilon/ d\eta=\partial\epsilon/\partial\eta=0\Rightarrow \epsilon=E(y^a)$.\

\section{The Jacobi Metric and the Lyapunov exponent}

Being $\epsilon$ a constant of motion, the term $\epsilon \partial_\eta \tau=E(y^a)\partial_\eta \tau$ in (\ref{hamiltoniana ridotta}) behaves as an exact time derivative; hence the variational principle rewrites: 
\begin{equation}
\label{bo}
	\delta\int d^3y(p_\xi d\xi+p_\theta d\theta)=0\, ,
\end{equation}
coupled with the constraint (\ref{hamiltoniana ADM}).\
This dynamical scheme allows to construct the Hamilton-Jacobi metric \cite{A} corresponding to the dynamical flow. Indeed for each point of the space it can be reproduced the same analysis developed in \cite{IM01} for the homogeneous Mixmaster model; in particular all the spatial gradients are dumped and the space points dynamically decouple in the asymptotic limit to the singularity.\
In fact, in each space point, the system dynamics is replaced by a geodesic flow $\delta\int ds=0$, with 
\begin{equation}
\label{elemento di linea}
	ds^2=E^{2}(y^a)\left(\frac{d\xi^{2}}{ (\xi^{2}-1)}+(\xi^{2}-1)d\theta^{2}\right)\,,
\end{equation}
corresponding to the Jacobi line element.The Jacobi metric is valid independently in each of such point (here the space coordinates behaves like external parameter) since the evolution is spatially uncorrelated.
The Ricci scalar takes the value $R=-2/E^2$, hence such a metric describes a two-dimensional Lobachevsky space; the role of the potential wall (\ref{U}) consists of cutting a closed domain $\Gamma_Q$ on such a negative curved surface.  Thus, summarizing, the system obtained is isomorphic to a billiard on a Lobachevsky plane.\\

A precise information about the dynamical stability of the geodesic flow associated with the line element (\ref{elemento di linea}) arises by the calculation of the Lyapunov exponent.\
Let us project the connecting vector $Z^\mu$ between two close geodesics on a Fermi basis $\{u^\mu, w^\mu\}, (\mu=1,2)$; the geodesic vector $u^\mu$ is taken in the form
\begin{equation}
	\label{jacobi quadrivelocita cap5}
u^{\mu}=\left(\frac{d\xi}{ ds},\frac{d\theta}{ ds}\right)=\left(\frac{\sqrt{\xi^{2}-1}}{
E}\cos\phi(s),\frac{\sin\phi(s)}{ E \sqrt{\xi^{2}-1}}\right)
\end{equation}
where $s$ denotes the curvilinear coordinate, while $\phi(s)$ is an angular variable ($0\le \phi<2\pi$), whose dynamics is obtained by requiring that the geodesic equation is verified, i.e.
\begin{equation}
\label{equazione per phi}
	\frac{d\phi(s)}{	ds}=-\frac{\xi}{ E\sqrt{\xi^{2}-1}}\sin\phi(s)\, .
\end{equation}
The vector $w^\mu$ is determined by the  property of the Fermi basis to be orthonormal, and it reads explicitly
\begin{equation}
\label{quadrivettore ortonormale}
w^\mu=\left(-\frac{\sqrt{\xi^2-1}} {E} \sin\phi; \frac{\cos\phi}{E\sqrt{\xi^2-1}}\right)\, .
\end{equation}
Projecting the connecting vector $Z^\mu$ over the Fermi basis defined above
\begin{equation}
	Z^\mu=Z_u(s)u^\mu+Z_w(s) w^\mu\, ,
\end{equation}
from the geodesic deviation equation we get that the dynamics is described by the following equations\
\begin{equation}
\label{sistema}
\begin {cases}
	\frac{d^2Z_u}{ ds^2}=0\cr
	
	\frac{d^2Z_w}{ ds^2}=\frac{Z_w}{ E^2}
	\end{cases}
	\end{equation}
The solution for the system (\ref{sistema}) is given by
\begin{equation}
\label{soluzioni}
	\begin{cases}
	Z_u=As+B,\ \ \ A,B=\textrm{const}\cr
	Z_w=c_1 e^{s/E}+c_{2}e^{-s/E},\ \ \ c_1,c_2=\textrm{const}\cr
	\end{cases}
\end{equation}
The value of $E$ given by the constraint (\ref{hamiltoniana ADM})and involved in the line element (\ref{elemento di linea})is determined by the initial conditions and cannot vanishes.\
By the first of solutions (\ref{soluzioni}) no geodesic deviation takes place along the geodesic vector (as expected); instead from the second solution we get a non-zero Lyapunov exponent of the form
\begin{equation}
\label {Lyapunov}
	\lambda(y^a)=\limsup_{s\to\infty}\frac{\ln(Z_w^{2}+(d Z_w/ds)^{2})}{ 2s}=\frac{1}{ E(y^a)}>0\,.
\end{equation}
For the validity of this analysis we have to verify that in the limit to the initial singularity the curvilinear coordinate $s$ approaches infinity; repeating the same procedure of \cite{IM01}, formulas (31) appearing there, in our inhomogeneous case is replaced by
\begin{equation}
	\frac{\partial s}{\partial\eta}=E(y^a)\partial_\eta \tau
\end{equation}
 i.e.(being $f(y^a)$ a generic function of the space coordinates)
\begin{equation}
\label{tempo}
	s=E(y^a)\tau+f(y^a)\Rightarrow \lim_{\tau\rightarrow\infty}s=\infty\, ;
\end{equation}
this ensures that the curvilinear coordinate $s$ behaves in the appropriate manner in the limit toward the singularity ($\tau\rightarrow\infty$).

Hence the chaoticity of the inhomogeneous Mixmaster dynamics is ensured by $\Gamma_Q$ to be a closed domain \footnote{Indeed the compact phase space is constituted by $\Gamma_Q\otimes S^1_\phi$, being $S^1_\phi$ the unit $\phi$-circle}; the covariance of such a description follows from the  independence of the Lyapunov exponent with respect to the lapse function and the shift vector. In fact, in (\ref{condizione di gauge sullo shift-vector}) and (\ref{condizione di gauge sulla lapse function}) $N$ and $N_\alpha$ are fixed (in turn) by choosing the form of the quantities $y^a$ and the latter can be generic functions subjected only to the condition to be smooth enough.
Eq.(\ref{Lyapunov}) provides the form of the Lyapunov exponent in the whole space domain but we stress how its value depends on the choice of Misner-Chitr\'e like variables; the independence of our scheme on the shift vector is ensured by the asymptotic behavior of the potential term, but to get $\epsilon$ as constant of motion, allowing the Jacobi metric representation, we need Misner-Chitr\'e like variable.

The covariance of our picture is equivalent to the covariance of the inhomogeneous Mixmaster chaos because it is well known \cite{IM01,KM97P} that the obtained billiard has stochastic properties (see also \cite{B82}). In fact the negative curvature of the Lobachevsky plane makes unstable the geodesic flow; the potential walls have the role of replacing a given geodesic with a different one(whose tangent vector is related to the previous one by a reflection rule \cite{B82})
and as we will show in Section 6 their structure will influence the chaotic properties of the system dynamics.

To better characterize the chaoticity of the obtained billiard, we show that in each point of the space our system admits an invariant measure which in the present variable is uniform over the admissible phase space.

From the point of view of statistical mechanics, such a system 
admits, point by point in space, an ``energy-like'' constant of motion which corresponds
to the kinetic part of the ADM Hamiltonian $\epsilon = E(y^a)$. 
The point-universe, randomizing within the closed domain $\Gamma_Q$, is represented by a dynamics which allows for 
an ensemble representation; in view of the existence of the ``energy-like''
constant of motion, the system evolution is appropriately described by {\it microcanonical ensemble}.
Therefore the stochasticity of this system is governed by the
Liouville invariant measure
\begin{equation} 
d\rho \propto  \delta \left(E(y^a) - \epsilon \right)d\xi d\theta dp_{\xi }dp_{\theta }  
\label{u} 
\end{equation} 
where $\delta\left(x\right)$ denotes the Dirac functional.

Since the particular value taken by the function 
$\epsilon$ $(\epsilon=E(y^a))$ cannot influence 
the stochastic property of the system
and must be fixed by the initial conditions, 
then we must integrate (in functional sense) over all admissible form of $\epsilon$. To do this 
it is convenient to introduce the natural variables 
$(\epsilon,\varphi)$ in place of $(p_\xi,p_\theta)$ by
\begin{eqnarray} 
p_{\xi } &=& \frac{\epsilon}{\sqrt{\xi ^2 - 1}}\cos\varphi \nonumber \\ 
p_{\theta } &=& \epsilon \sqrt{\xi ^2 - 1}\sin\varphi ,  
\label{v} 
\end{eqnarray} 
where $0 \leq \varphi < 2\pi$.
By integrating over all 
functional forms of $\epsilon$,we removes the Dirac delta functional, 
which leads in each point of space to the uniform normalized invariant measure 
\footnote{Since the space points are dynamically decoupled the whole invariant measure of the system would correspond to the infinite product of (\ref{x}) through the space domain}

\begin{equation} 
d\mu(y^a) = d\xi d\theta d\varphi \frac{1}{8\pi ^2} . 
\label{x} 
\end{equation}

The existence of the above stationary probability distribution in $\Gamma_Q$, outlines the chaotic properties associated to the point-like billiard out coming from our analysis. 

As we stressed, our dynamical scheme relies on the use of Misner-Chitr\'e like variables and therefore the covariance of the Lyapunov exponent is invariant with respect to space-time coordinates, but it could be sensitive to the choice of configurational variables. Thus the result here obtained calls attention to be extended to any choice of the configurational variables.
In this respect we compare our result with the analysis presented in \cite{M03} according to which, given a dynamical system of the form
\begin{equation}
	d{\bf x}/dt ={\bf F}(x)
\end{equation}
then the positiveness of the associated Lyapunov exponents are invariant under the following diffeomorphism: ${\bf y}=\phi({\bf x},t), d\tau=\lambda({\bf x},t)dt$, as soon as the four requirements hold:

\begin{enumerate}
{\item the system is autonomous}
{\item the relevant part of the phase space is bounded}
{\item the invariant measure is normalizable}
{\item the domain of the time parameter is infinite}
\end{enumerate}

To show that such a covariance criterion is here fulfilled, we observe that the variables ${\bf x}$ can be identified with $\tau,\xi,\theta$ and $\varphi$,and the time variable with our curvilinear coordinate $s$. On the other hand the above diffeomorphism relation in its time independent-form can match a phase-space coordinates transformation; then we underline also that:
\begin{enumerate}
{\item in the considered asymptotic limit our dynamical system is autonomous because its hamiltonian coincides with the constant of motion  $\epsilon=E(y^a)$ and the potential walls are fixed in time.}
{\item apart from sets of zero-measure which cannot be explored by the system \cite{BKL70}, the phase-space $\Gamma_Q$ is a compact domain. }
{\item as shown by the above analysis which leads to (\ref{x}) the system admits, in each space point, a normalized invariant measure over the phase space.}
{\item we showed by (\ref{tempo}) that the curvilinear coordinate $s$ admits an infinite domain because the variables $\tau\in(-\infty,\infty)$. }
\end{enumerate}

Thus, on the base of \cite{M03},we can claim that the Lyapunov exponent calculated in (\ref{Lyapunov}) provide an appropriate chaos indicator only when the effects of the boundary are taken into account in agreement with our discussion of the next section. Furthermore such an indicator is covariant with respect  any configurational coordinate transformation which preserves the requirements 1-4.
Strikingly, we have to stress that if we adopt Misner like variables \cite{M69} the Lyapunov exponent (\ref{Lyapunov}) is no longer a good indicator; in fact the anisotropy parameters (\ref{anisotropia}) in Misner-like variables depend not only on $\beta_+,\beta_-$, but also on $\alpha$, and therefore after the ADM reduction on the curvilinear coordinate $s$. As a consequence, the conditions 1 and 3 are no longer full filled for this choice because the potential walls move with time.
However for any generic transformation of coordinates which involves only $\xi$ and $\theta$, the chaoticity of the  Mixmaster model is preserved.
We conclude this section by observing that if in different configurational coordinates the inhomogeneous Mixmaster would not appear as chaotic, then the stochasticity has to be transferred to the coordinate transformation which link them to the Misner-Chitr\'e like variable.

\section{The role of the potential walls}

Up to now we involve the role of the potential walls in order to cut a billiard on the Lobachevsky plane. Here we discuss a notion of Lyapunov exponents which include either the feature of the geodesic flow either the structure of the bounding potential walls, and we arrive to show that even in this more complete framework our system is a chaotic one.
To this end we are showing below that our system meets all the hypothesis at the ground of the Wojtkowski Theorem, see \cite{woj}.
Let us consider the following new choice of coordinates on the 2-surface:

\begin{equation}
\label{y xi}
\vec{y}=(y_1,y_2)=\frac{(1+\xi)}{\sqrt{\xi^2-1}}(\cos\theta,\sin\theta)
\end{equation}

On the basis of (\ref{y xi}) the new line element and the anisotropy parameters read as
\begin{equation}
\label{line element}
ds^2={\frac{4E^2 d\vec y^2}{(1-y^2)^2}};\,\,\,\, y<1;
\end{equation}

\begin{equation}
Q_a=[(\vec y ^2+A_a)^2+1-(A_a)^2];\,\,\,a=1,2,3
\end{equation}
being $A^1_a(-\sqrt3,\sqrt3,0)$ and $A^2_a(1,1,-2)$

Since here we introduce the Poincar\'e model of the Lobachevsky plane in the form of the upper-half plane; this is reached by using the well-known Poincar\'e variables

\begin{equation}
\vec\eta=2{\frac{\vec y + \vec b}{(\vec y +\vec b)^2}}-\vec b
\end{equation}

where $\vec b$ denotes a point on the absolute ($b^2=1$). In terms of these new variables the metric (\ref{y xi}) takes the form
\begin{equation}
ds^2={\frac{(d\vec\eta)^2}{(\vec\eta \cdot\vec b)^2}}
\end{equation}

and the available domain $\left| y\right|\le1$ transforms into the half-plane $(\vec\eta \cdot\vec b)\ge0$ while the absolute represents the line $(\vec\eta \cdot\vec b)=0$. Now if we write
\begin{equation}
\vec\eta=\frac{2}{\sqrt3}\left[\left(u+\frac{1}{2}\right)\vec b^\perp+v \vec b\right] v\ge0
\end{equation}
where $\vec{b}=(0,1),\vec{b}^\perp=(1,0)$, then is easy to verify that in term of these coordinates the anisotropy functions have the form
\begin{equation}
	\label{anisotropie geodetiche}
	\begin{cases}
	Q_1(u,v)={\frac{-u}{u^2+u+1+v^2}}\cr
	Q_2(u,v)={\frac{1+u}{u^2+u+1+v^2}}\cr
	Q_3(u,v)={\frac{u(u+1)+v^2}{u^2+u+1+v^2}}
	\end{cases} 
\end{equation}

This is a very suitable expression for the boundary; in fact geodesic on this half-plane are semicircles having centers on the absolute (i.e. $u(s)=A+R\cos s, v(s)=R\sin s$ being $(A,0)$ the coordinate of the semicircle center and $R$ the corresponding radius) and rays being perpendicular to the absolute; the billiard is bounded by the geodesic triangle $u=0, u=-1$, and $(u+1/2)^2+v^2=1/4$. The new domain is shown in fig 2.

\begin{figure} 

\begin{center} 

\includegraphics[width=\hsize]{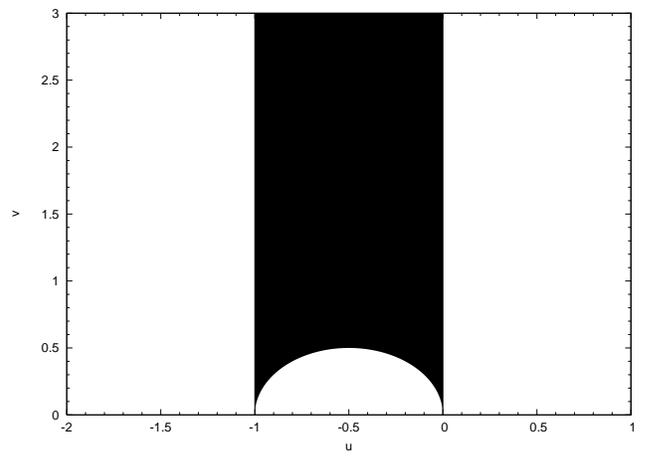}

\caption{ $\Gamma_Q(u,v)$\label{fig:cuspidi2 cap51}} 


\end{center}

\end{figure}

The billiard has a finite measure, and its open region at infinity together with the 2 points on the absolute $(0,0)$ and $(-1,0)$ correspond to the three cuspids of the potential in figure 1.
The tangent field to the geodesic flow takes the explicit form:
\begin{equation}
t^\mu\equiv(-v,u-A)
\end{equation}
and the associated matrix of the dynamics $M_\mu^\nu\equiv\partial_\mu t^\nu$ is constant and orthonormal.

Referring to the notation of \cite{woj} we can easily construct in this variables the cocycle  of the dynamics and verify that theorem 2.2 (pag 151 of this reference) applies; since the manifold is connected, has finite volume, and the matrix of the cocycle is constant, we have only to find an invariant bundle of sectors.
This is possible because the following two properties hold 
\begin{enumerate}
{\item because of the constant negative curvature of the surface, a one-parameter family of geodesics with negative curvature is invariant under the evolution}
{\item the bounding potential walls are constituted by two straight lines and a semicircle of negative curvature; the former ones do not affect the structure of the cones during the bounces of the geodesic, while the latter one, being a dispersing profile, ensures that after reflection against it, the cones will evolve in themselves.}
\end{enumerate}

After this discussion we can claim that the largest Lyapunov exponent has positive sign almost everywhere.
The covariance of this result is ensured in view of the discussion developed in the previous section based on (\cite{M03}). The analysis of this section complete our analysis about the covariant nature of the chaos associated to the billiard representation of the Mixmaster model.

\section{Concluding remarks}

The main issue of the present work consists of the proof that the chaotic behavior singled out by a generic inhomogeneous model near a singularity has a covariant nature. This result has been obtained by a "gauge" independent ADM reduction of the dynamics to the physical degrees of freedom, which for the Universe correspond to the anisotropy degrees, i.e. to the functions $\xi$ and $\theta$.\\

We describe the evolution as independent of the lapse function and the shift vector form by adopting the variables $y^a$ as the new spatial coordinates.\
However their degrees of freedom do not disappear from the problem because they are transferred to the SO(3) matrices $O^a_b$ which acquire a dependence on time in the new variables; such a dependence is then eliminated from the dynamics by solving the super-momentum constraint and using some implications deriving by the approximation of $\sqrt {g} ^{(3)}R$ as an infinite potential wall.\
The potential behavior (\ref{U}) is crucial for the existence of an "energy-like" constant of motion $\epsilon\equiv E(y^a)$, and therefore is at the basis of the chaos description.\
In this view, this approximation is naturally induced for $D\to 0$, due to the potential structure; the only assumption required is for the functions $y^a$ and $O^a_b(y^c)$ to be smooth ones, in the sense that their presence does not affect the asymptotic behavior of the potential term. This restriction is a natural one having a good degree of generality, because the smoothness of the functions $y^a$ is ensured by the smoothness of the lapse function and the shift vector (see (\ref{condizione di gauge sullo shift-vector}-\ref{condizione di gauge sulla lapse function})), i.e. by the choice of a regular reference frame. The initial smoothness of the matrices $O^a_b$ (when are taken on the spatial coordinates $x^a$) is preserved by the coordinates transformation.\
Concluding, the cosmological meaning of this concept corresponds to deal with independent "horizons"; the approximation neglecting in (\ref{hamiltoniana ADM}) the potential term with respect to the value of $"\epsilon"$ is equivalent to require the scale of the inhomogeneities to be super-horizon sized (see \cite{BKL82, K93, M95}).

\section{Acknowledgment}

Carlangelo Liverani is thanked for his very valuable comments on this subject.






\begin{thebibliography}{99}
%
%
\bibitem{KT90} E.W. Kolb and M.S. Turner, \textit{The Early Universe},
(Adison-Wesley, Reading) (1990).

\bibitem{deB00}
P. De Bernardis et al., \textit{Nature}, \textbf{404}, 955 (2000).

\bibitem{MAP}
D.N. Spergel et al., \textit{Astrophys. J. Suppl. Serv.}, \textbf{148}, 175, (2003).

\bibitem{BKL82}  V.A. Belinskii, I.M. Khalatnikov and E.M. Lifshitz, \textit{Adv. Phys.}  \textbf{31},  639 (1982).

\bibitem{K93}
A.A. Kirillov,
{\it Zh.\ Eksp.\ Theor.\ Fiz.}, {\bf 103}, 721-729, (1993).

\bibitem{M95}
G. Montani, \textit{Class. and Quantum Grav.}, \textbf{12}, 2505, (1995).

\bibitem{KL63} E.M. Lifshitz and I.M. Khalatnikov, \textit{Adv.Phys.}, \textbf{12}, 185 (1963).

\bibitem{LL} L. L. Landau and E. M. Lifshitz, {\it Fields Theory}, Mir (1985).

\bibitem{GRI}
L.P.Grishchuk, A.G.Doroshkevich and V.M.Yudin
{\it Zh. Eksp. Teor. Fiz.},{\bf 69} (1975),1857.

\bibitem{HH83} 
J.B. Hartle and S.W. Hawking, 
{\it Phys. Rev. D},  {\bf 28}, 12, (1983). 

\bibitem{KM97J} 
A.A. Kirillov and G. Montani, 
{\it JETP Lett.}, {\bf 66}, n 7, 475, (1997). 


\bibitem{KM02} A.A. Kirillov and G. Montani, {\it Phys. Rev. D}, {\bf 66}, 064010 (2002).

\bibitem{BKL70} 
V.A. Belinski, I.M. Khalatnikov and E.M. Lifshitz, {\it Adv.\ Phys.}, {\bf 19}, 525, (1970). 


\bibitem{M03A}
G.Montani,
{\it Int. Journ. Mod. Phys. D},(2003),{\bf 12},No 8, 1445.


\bibitem{M69} 
C.W. Misner, {\it  Phys.\ Rev.\ Lett.}, {\bf 22}, 1071, (1969). 

\bibitem{WIB}
M.Weaver,J.Isenberg, B.K.Berger,
{\it Phys. Rev. Lett.}{\bf 80} 2984-2987, (1998)

\bibitem{BERGER}
B.K.Berger,
{\it Proceedings of an International Conference held at the University of Alabama in Birmingham}, edited by R.Weillard,G.Weinstein (American Mathematical Society, 2000)


\bibitem{M01}
A.E.Motter,P.S.Letelier,
{\it Phys. Lett. A}{\bf 285} 127-131, (2001)

\bibitem{M03}
A.E.Motter,
{\it Phys. Rev. Lett.}{\bf 91} 231101, (2003)


\bibitem{CL97} 
N.J. Cornish, J.J. Levin, 
{\it Phys. Rev. Lett.}, {\bf 78}, 998, (1997); {\it Phys.\ Rev.\ D}, 
{\bf 55}, 7489, (1997). 



\bibitem{H94} 
D. Hobill, A. Burd, A. Coley, (eds) {\it Deterministic Chaos in General Relativity}, (World Scientific, Singapore) 
(1994). 



\bibitem{IM01}
G.P. Imponente and G. Montani, {\it Phys. Rev. D}, {\bf 63}, 103501, (2001).

\bibitem{KM97P} 
A.A. Kirillov and G. Montani, {\it Phys.\ Rev.\ D}, {\bf 56}, n. 10, 6225, (1997). 

\bibitem{CB83} 
D.F. Chernoff and J.D. Barrow, {\it Phys.\ Rev.\ Lett}, {\bf 50}, 134,(1983). 

\bibitem{M00}  G. Montani, \textit{Class. and Quantum Grav.},  
\textbf{17}, 2205 (2000).

\bibitem{C72} 
D.M. Chitr\'e, 
{\it Ph.D. Thesis}, University of Maryland, (1972). 

\bibitem {A}
V.I. Arnold, {\it Mathematical Methods of Classical Mechanics}, Springer-Verlag (Berlino,1989).

\bibitem{B82} 
J. D. Barrow  
{\it Phys.\ Rep.}, {\bf 85},n. 1,1-49 (1982).

\bibitem{woj}
M. Wojtkowski,
{\it Ergod. Th. and Dynam. Sys.},{\bf 5},145-161 (1985)











\end{thebibliography}
\end{document}